\documentclass[12pt]{article}
\usepackage{epsfig}
\begin{document}
\title{A High Current Proton Linac with 352 MHz SC Cavities}
\author{C. Pagani, G. Bellomo, P. Pierini \\
INFN - Sezione di Milano - Laboratorio LASA, \\
Via Fratelli Cervi 201, 
20090 Segrate (MI) Italy\\[0.3cm]
{\it\small
To Appear in the Proceedings of the 1996 LINAC Conference,}\\
{\it\small Gen\`eve August 26-30 1996}}
\date{ }
\maketitle
\begin{abstract}
A proposal for a 10-120 mA proton linac
employing superconducting beta-graded, CERN type, four cell
cavities at 352 MHz is presented.

The high energy part (100 MeV-1 GeV) of the machine is split 
in three $\beta$-graded sections, and transverse focusing 
is provided via a periodic doublet array.
All the parameters, like power in the couplers and accelerating 
fields in the cavities, are within the state of the art, 
achieved in operating machines.

A first stage of operation at 30 mA beam current is proposed, 
while the upgrade of the machine to 120 mA operation  
can be obtained increasing the number 
of klystrons and couplers per cavity. The additional coupler 
ports, up to four, will be integrated in the cavity design. 
Preliminary calculations indicate that beam transport
is feasible, given the wide aperture of the 352 MHz structures.

A capital cost of less than 100 M\$ at 10 mA, reaching up 
to 280 M\$ for the 120 mA extension, has been estimated for 
the superconducting high energy section (100 MeV-1 GeV).

The high efficiency of the proposed machine, reaching 50\%
at 15 mA, makes it a good candidate 
for proposed nuclear waste incineration facilities
and Energy Amplifier studies\cite{proposta,lanl-acc}.
\end{abstract}

\section{Choice of the 352 MHz Frequency}

Our design is based mainly on the choice of a low RF 
frequency for the SC linac. 
A wide experience in the design, construction and operation of
352 MHz cavities and RF systems is available at CERN\cite{cern}.
The 352 MHz frequency at moderate
gradient operation (around 5 MV/m) allows for large 
geometrical irises and lower beam current densities.
A critical issue for such a machine will be the 
control of the beam halo growth\cite{lanlhalo}, 
and the choice of a low frequency allows to lower both the 
space charge tune depression and the ratio of the 
beam (core) size with respect to the beam line aperture.

Another important issue, the future availability 
of several 1.3 MW CW klystrons of the CERN LEP RF system, 
that will be decommissioned before year 2000,
gives an economical impulse for the investigation of 
a scheme based on the LEP 352 MHz frequency. 
Moreover, we have also to take into account the experience of 
several European companies for cavities 
production and the cavity tooling machines already 
available at the companies\cite{cern}.

In our view the developement of new $\beta$-graded 
structures\cite{argonne,lanl-acc}, with up to four coupler ports, 
at 352 MHz could allow to reach a 
beam current of 120 mA employing present technological RF 
components (simply by incrementing the number of klystrons 
and couplers/cavity, limiting the power per coupler to approximately 
200 kW).

In the following we present a preliminary parameter 
set for the high energy part (100 MeV-1 GeV) of 
the machine, as 
presented to C.~Rubbia in the framework of a possible INFN 
collaboration to the Energy Amplifier and waste transmutation project.
The low energy part should be composed of two sections:
an RFQ\cite{lanl-acc} (up to $\approx$ 7 MeV) and 
a conventional DTL linac (up to $\approx$ 100 MeV).
This design has been recently included as the
candidate for the high energy accelerator section 
of the Energy Amplifier proposal\cite{rubbia2}, 
and work is in progress for a full optimization of 
the optics and for the development of the RF cavities.

\section{The $\beta$-graded structures for the high energy section}

We have chosen to cover the energy range from 
100 MeV to 1 GeV with three different families of $\beta$-graded 
four cell cavities at 352 MHz, with cell length defined as
$L_{\rm cell}={\bar\beta \lambda_{\rm RF}}/{2}$.
Four cell cavities have been chosen in order to reduce the 
number of cavities and the
physical structure length, that has to include cutoff tubes,
coupler and HOM ports.

This choice of three energy ranges (and consequently of
three $\bar\beta$ values for the different sections) 
allows to keep the transit time
factor of a particle in each cavity 
always greater than 0.9, along the whole machine.

The main characteristics of the cavities in each section
are given in Table~\ref{tabellabeta}.

\begin{table}[htb]
\begin{center}
\begin{tabular}{llll}
\hline\hline
Energy (MeV)& $\bar\beta$ &  $L_{\rm active}$ (m) &  $L_{\rm FIDO}$ (m) \\
\hline
100--185  & 0.47 & 0.800 & 7.5 \\
185--360  & 0.60 & 1.022 & 8.4 \\
360--1000 & 0.76 & 1.294 & 9.5 \\
\hline\hline
\end{tabular}
\end{center}
\caption{Energy range, design $\bar\beta$, 
active length and length of the focussing period, for the
three families of 4 cell cavities.}
\label{tabellabeta}
\end{table}

The energy gain in each cavity is given by:
\[
\Delta T_{\rm cav}(MeV)=L_{\rm active}(m) E_{\rm acc}(MV/m) 
      g\left(\frac{\bar\beta}{\beta}\right)
      \cos\left(\phi_{\rm RF}\right)
\]
where $L_{\rm active}=N L_{\rm cell}$ is the active cavity length 
(in Table~\ref{tabellabeta}),
$E_{\rm acc}$ is the accelerating field in the cavity,
$g(\bar\beta/\beta)$ is the transit time factor of the cavity,
depending on the design $\bar\beta$ and the actual beam $\beta$, 
and $\phi_{\rm RF}$ is the operating RF phase.

\begin{figure}[hbt]
\begin{center}
\epsfig{file=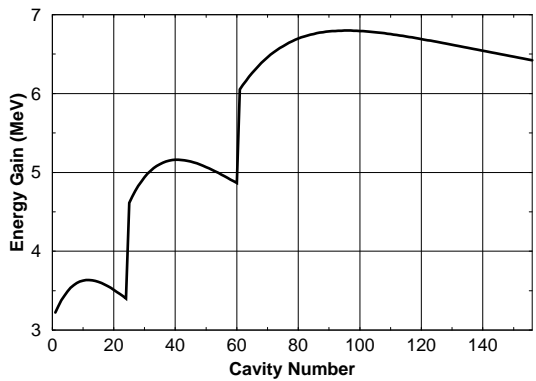,bbllx= 45pt,bblly=50pt,
                       bburx=520pt,bbury=380pt,width=7.5cm,clip=}
\end{center}
\caption{Energy gain along the three linac sections, as a function of 
the cavity number, keeping the nominal accelerating 
gradient fixed in each section (see text for details).}
\label{egain}
\end{figure}

In our design, considering the requirement of phase stability, 
we have chosen $\phi_{\rm RF}=-30^\circ$, and $g>0.9$ all over 
the machine.
The role of the transit time factor $g$ can be seen
in Figure~\ref{egain}, where we plot the 
energy gain of each cavity along the machine. Here we
chose a constant $E_{\rm acc}$ in each section 
(the values are given in Table~\ref{sezioni}); as an 
alternative approach one can individually set the cavity 
gradients to provide a constant energy gain in the sections.

The basic accelerating cell of each linac section 
consists of one cryomodule containing four cavities, 
transverse focusing is provided by quadrupole 
doublets every cryomodule. 

\section{Focusing Structure}

\begin{figure}[htb]
\begin{center}
\epsfig{file=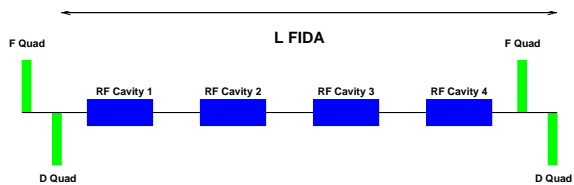,bbllx= 0pt,bblly=0pt,
                            bburx=179pt,bbury=586pt,
                            angle=-90,width=7.5cm,clip=}
\end{center}
\caption{Focusing structure of the linac.}
\label{fido}
\end{figure}

The focusing structure is a FIDA cell, 
where the beam acceleration is provided 
by four RF cavities, in one cryomodule,
between successive quadrupole
doublets, as seen in Figure~\ref{fido}. 
The possible use of quadrupole triplets to allow for 
``rounder'' beams will also be considered.

The three sections of the linac have different cell lengths.
The active cavity length 
and the corresponding lattice periodicity 
in the three sections are indicated in Table~\ref{tabellabeta}.

The quadrupole integrated field $Gl$ ranges from 1 to 3.5 T 
along the machine, hence it is possible to place 
warm normal conducting quadrupoles 
between the cryomodules.

For this reference design the zero current maximal 
phase advance per cell has been set to
90$^\circ$, although a value close to 60$^\circ$ 
or 72$^\circ$ should be more appropriate. In the first unit 
of the first section there is a strong longitudinal phase 
advance and a reduction of $E_{\rm acc}$ will be investigated.

Preliminary calculations with the linear space charge 
code TRACE-3D\cite{trace3d} in the current
range of 10-120 mA show that beam transport with a 
beam radius/aperture ratio in the range 10--50 along the machine is
possible. 

\section{Section details}

In Table~\ref{sezioni} we report the main characteristics of the
three sections of the high energy part of the linac, including 
RF power distribution. 
The maximum RF power in the couplers
is approximately 200 kW, 
the current upgrade would require the insertion of additional 
couplers (up to four) in each cavity. The four coupler ports 
should be integrated from the beginning in the cavity design.

\begin{table}[htb]
\begin{center}
\begin{tabular}{lccc}
\hline\hline
                      &  S. 1 &  S. 2  &  S. 3 \\
\hline
 N. of structures     &  24       &   36       &   96     \\
 $\bar\beta$          & 0.47      & 0.60       & 0.76     \\
 $E_{\rm acc}$ (MV/m) & 5.2  & 5.8        & 6.0      \\
 Section length (m)   & 45     & 76         & 226      \\
\hline
\multicolumn{4}{c}{ 10 mA beam current}\\
\hline
 RF Power/section (MW) & 0.85 & 1.81  & 6.35 \\
 RF Power/cavity (kW)  & 35.4 & 50.3 & 66.1 \\
 couplers/cavity       & 1 & 1 & 1 \\
 Klystron/section      & 1 & 2 & 6 \\
\hline
\multicolumn{4}{c}{ 120 mA beam current}\\
\hline
 RF Power/section (MW) &  10.2 &  21.72  &  76.2 \\
 RF Power/coupler (kW) & 212   & 201     & 198 \\
 couplers/cavity       &   2   &   3     &   4 \\
 Klystron/section      & $\approx$10 & $\approx$20 & $\approx$80 \\
\hline\hline
\end{tabular}
\end{center}
\caption{Section details.}
\label{sezioni}
\end{table}

\subsection{Estimated RF Capital Cost}
A total of 156 cavities and 350 m of physical length 
are required for the three sections of the superconducting
linac, in this reference design. These two numbers could slightly
increase in the final design, in order to: decrease the cavity 
gradient, employ quadrupole triplets focusing, include 
beam diagnostic elements inside the cryomodules
or matching elements between sections.

\section{Cost of the linac, and efficiency considerations}

The capital cost of the superconducting linac, excluding the
RF power costs, is approximately 72.5 M\$, and the cost breakdown is 
indicated in Table~\ref{capitalcost}.

\begin{table}[htb]
\begin{center}
\begin{tabular}{lcr}
\hline\hline
 Item                           &  Number &  M\$ \\
\hline
Cavities (with tuners)          & 156    & 39.0 \\
Quadrupoles                     &  78    &  3.1 \\
RF Controls                     & 156    &  4.2 \\
Vacuum Pums                     &  40    &  1.0 \\
Vacuum Valves                   &  78    &  1.0 \\
Cryostats                       &  39    &  8.0 \\
Beam Monitors                   &  80    &  0.8 \\
Controls                        &   1    &  3.0 \\
Cryoplant Cost (8 kW @ 4.2 K)   &   1    &  8.9 \\
Ancillary Equip. (350 m)        &        &  3.5 \\
\hline
 Total Cost                     &        & 72.5 \\
\hline\hline
\end{tabular}
\end{center}
\caption{Capital cost of the SC linac.}
\label{capitalcost}
\end{table}

\begin{figure}[htb]
\begin{center}
\epsfig{file=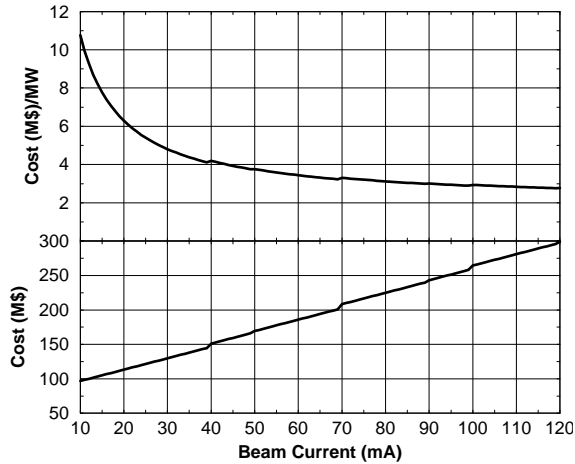,bbllx= 45pt,bblly= 50pt,
                       bburx=520pt,bbury=450pt,width=7.5cm,clip=}
\end{center}
\caption{Lower curve: Linac total capital cost (in M\$, including RF system).
Upper curve: Total cost per MW of beam power vs. beam current (mA).}
\label{f:totalcost}
\end{figure}

Assuming, 1.5 M\$ per klystron (1.3 MW, CW with power supplies) and
50 k\$ per coupler (including the RF distribution), in 
Figure~\ref{f:totalcost} we plot the total 
capital cost of the linac (including the RF system), and the 
total capital cost per MW of beam power,
as a function of the beam current (in mA).

\subsection{Overall Linac Efficiency vs. Beam Current}

\begin{figure}[htb]
\begin{center}
\epsfig{file=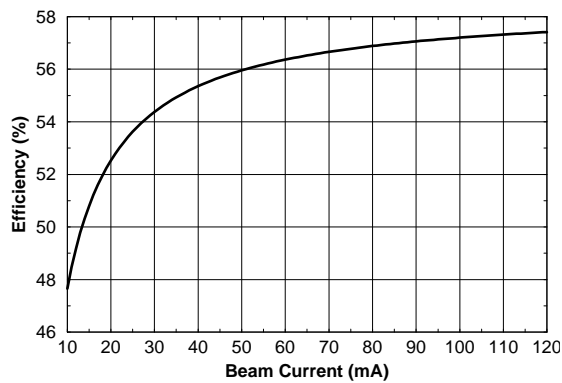,bbllx= 45pt,bblly= 50pt,
                     bburx=520pt,bbury=380pt,width=7.5cm,clip=}
\end{center}
\caption{Overall linac plug efficiency vs. beam current (mA)}
\label{f:eff}
\end{figure}

Assuming a klystron efficiency of 58\%, a refrigeration power of 
approximately 2.5 MW and a contingency power of 1 MW dedicated to
the ancillary components of the linac, the overall efficiency 
of the machine as a function of the beam current is presented
in Figure~\ref{f:eff}. Note that 50\% plug efficiency is reached at
15 mA operation. The operation at the full 120 mA current would 
allow to reach nearly the nominal klystron efficiency.

\section{Conclusions}
A preliminary study for a low frequency, high current 
superconducting proton linac 
for nuclear waste incineration 
and energy amplifier 
applications has been proposed. 
The machine operates at the 352 MHz of the LEP RF system  
with three sections of $\beta$-graded superconducting cavities.

Preliminary calculations indicate that beam transport at 
high current is possible, and further studies to address 
cavity design, both from 
the electromagnetic and the engineering point of view, 
and beam halo formation are in the starting phase.

The choice of the RF frequency and of the machine parameters 
provides a very good plug efficiency at high beam current,
a crucial issue for the proposed applications.

\section{Acknowledgements}

We are grateful to Carlo Rubbia who stimulated this work.

\end{document}